\documentclass[10pt,twocolumn]{article}

\usepackage{graphicx}
\usepackage{amssymb}
\usepackage{enumerate}
\usepackage{color}

\newcommand{\be}{\begin{equation}}
\newcommand{\ee}{\end{equation}}
\newcommand{\n}{\noindent}
\newcommand{\dd}{\,{\rm d}}

\begin{document}




\title{Lipid Domain Order and the Algebra of Morphology}
\author{\small Tristan S. Ursell\,\,\,\,and Rob Phillips\footnote{Address correspondence to: phillips@pboc.caltech.edu}\\ 
\small Department of Applied Physics, California Institute of Technology, Pasadena, CA 91125\\}
\date{}
\maketitle


{\footnotesize
Lipid  membranes regulate the flow of materials and information between cells and their organelles.   Further, lipid composition and morphology can play a key role in regulating a variety of biological processes. For example, viral uptake, plasma membrane tension regulation, and the formation of caveolae all require the creation and control of groups of lipids that adopt specific morphologies.  In this paper, we use a simplified model mixture of lipids and cholesterol to examine the interplay between lipid phase-separation and bilayer morphology.  We observe and theoretically analyze three main features of phase-separated giant unilamellar vesicles.  First, by tracking the motion of `dimpled' domains, we measure repulsive, elastic interactions that create short--range translational and orientational order, leading to a stable distribution of domain sizes, and hence maintaining lateral heterogeneity on relatively short length scales and long time scales.  Second, we examine the transition to `budded' domain morphologies, showing that the transition is size-selective, and has two kinetic regimes, as revealed by a calculated phase diagram.  Finally, using observations of the interactions between dimpled and budded domains, we build a theoretical framework with an elastic model that maps the free energies and allowed transitions in domain morphology upon coalescence, to serve as an interpretive tool for understanding the algebra of domain morphology.  In all three cases, the two major factors that regulate domain morphology and morphological transitions are the domain size and membrane tension.
}\\



\begin{figure}
\begin{center}
\includegraphics[width=3.6in]{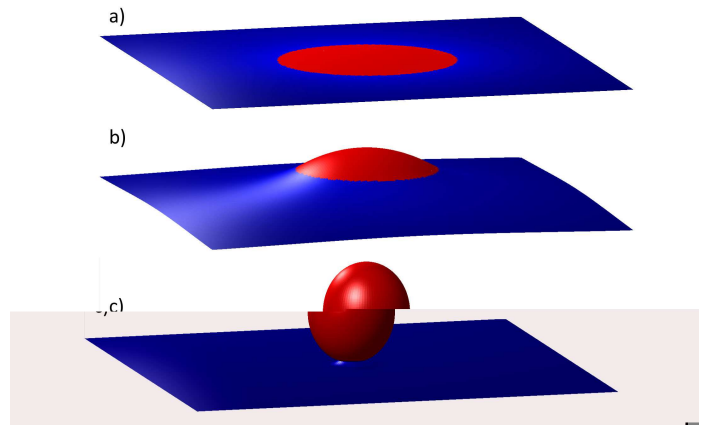}
\caption{\footnotesize{Morphologies of a lipid domain.  a)  A domain (red) lies completely flat when the energy from line tension is small compared to the cost of deformation from bending and membrane tension.  b) For domains with a size roughly equal to or less than the elastic decay length, a competition between bending and phase boundary line tension results in a morphological transition from a flat to a dimpled state.  This morphology facilitates elastic interactions between domains that slow the kinetics of coalescence significantly.  c)  Line tension in domains whose size is large compared to the elastic decay length, can cause a transition to a fully budded state.}}
\label{fig1}
\end{center}
\end{figure}

\begin{figure}
\begin{center}
\includegraphics[width=3.2in]{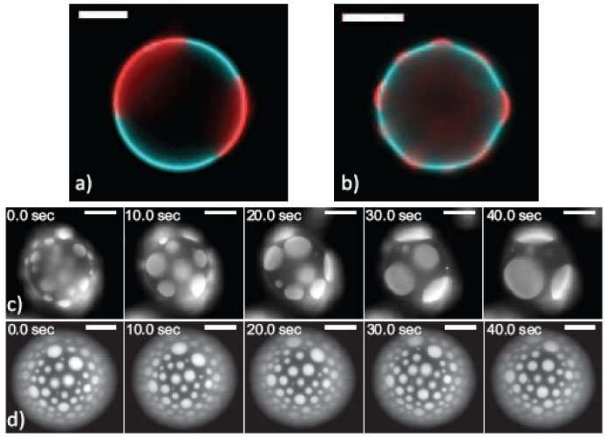}
\caption{\footnotesize{Domain morphology and coalescence.  a) A nearly fully phase-separated vesicle, showing domains (red) flat with respect to the background curvature of the vesicle (blue). b)  At low tension, domains (red) dimple and establish a non-zero boundary slope with respect to the curvature of the vesicle (blue).  c) Flat domains on the surface of a vesicle - coalescence is uninhibited by elastic interactions. d) Dimpled domains on the surface of vesicle - coalescence is inhibited by elastic interactions between the domains, and the domain-size distribution is stable. Directly measuring membrane tension disturbs the domain size evolution, however the magnitude of membrane fluctuations \cite{Yanagisawa2007,Bassereau2004} indicates that the tension in (c) is higher than the tension in (d). Scale bars are $10\,\mu\mbox{m}$.}}
\label{fig2}
\end{center}
\end{figure}

{\small
Cellular membranes are a complex mixture of lipids, membrane proteins, and small molecules ({\it e.g.} sterols) \cite{Singer1972,Simons1997}.  The membrane serves mainly as a chemical barrier and substrate for membrane proteins that are responsible for regulating the movement of materials and
information across the membrane.  However, there are a host of important tasks that require a change in membrane morphology, such as endo- and exo-cytosis \cite{Freund2005,Maxfield2004}, vesicular trafficking from the endoplasmic reticulum and Golgi apparatus to the plasma membrane \cite{JLS2007book}, and the regulation of tension in the plasma membrane \cite{SensPRE2006}.  While the role of proteins cannot be ignored in these instances ({\it e.g.} clathrin, COPI, COPII, caveolin, SNAREs, actin) \cite{Quest2004,Hinrichsen2006,Cai2007,Fix2004,McMahon2005,Antonny2006,Bonifacino2004,Merrifield2005,Tsujita2006,Yarar2005}, the lipid composition and bilayer morphology of the membrane play an important part \cite{JLS2007book,Maxfield1999,Maxfield2000,Simons1998,Simons2000,Sprong2004,Riezman2000}.  With that in mind, our goals in this paper are to examine how lipids in a model multi-component membrane spatially organize, how this organization relates to membrane morphology, or specifically membrane mechanics, and to examine how transitions in membrane morphology are regulated by bilayer mechanical properties.

{\it In vitro} studies have conclusively shown that lipids are capable of lateral self-organization \cite{Veatch2003,Baumgart2003,Shimshock1973}, facilitated by the structure of their hydrophobic regions and the presence of intercalated sterols.  Saturated lipids and cholesterol are sequestered from the membrane mix to form `lipid rafts' that serve as platforms for signaling and material transport across the membrane, with sizes ranging from $\sim50-500\,\mbox{nm}$ \cite{Simons1997,SensPRE2006,Simons2004,Schlegel1998,Chazal2003,RaoTraffic2004,Dietrich2001,Park1998,Helms2004,Lucero2004,Gaus2003}.  In addition to their unique chemical properties and protein-specific interactions, lipid rafts are mechanical entities in a thermal environment, and as such, our analysis focuses on continuum and statistical mechanics of phase-separated bilayers.



The remainder of the paper is organized as follows. The first section examines how dimpled domains, which exert repulsive forces on each other, spatially organize themselves into ordered structures as their areal density increases.  Specifically, we measure the potential of mean force and orientational order, finding that domains exhibit orientational and translational order on length scales much larger than the domains themselves.  The second section studies the transition to a spherical `budded' morphology \cite{Lipowsky1992,SensPRL2005}.  Using a mechanical model that combines effects from bending, line tension and membrane tension, we predict and observe size-selective budding transitions on the surface of giant unilamellar vesicles, and derive a phase diagram for the budding transition.  The final section considers the three lipid domain morphologies -- flat, dimpled and budded -- and constructs a set of transition rules that dictate the resultant morphology resulting from coalescence of two domains.  \\



\n
{\bf Spatial Organization of Dimpled Domains\\} 
Our analysis begins by viewing the bilayer as a mechanical entity, endowed with a resistance to bending \cite{Evans2000}, quantified by a bending modulus ($\kappa_b$) with units of energy; a resistance to stretch under applied membrane tension ($\tau$) \cite{Evans2000}, with units of energy per unit area; and in the case where more than one lipid phase is present, an energetic cost at the phase boundary, quantified by an energy per unit length ($\gamma$) \cite{Baumgart2007,Kuzmin2005}.  For a given domain size, the line tension between the two phases competes with the applied membrane tension and bending stiffness to yield morphologies that reduce the overall elastic free energy.  In particular, the bending stiffness and membrane tension both favor a flat domain morphology.  Conversely, the line tension prefers any morphology (in three dimensions) that reduces the phase boundary length.  A natural length-scale, over which perturbations in the membrane disappear, is established by the bending stiffness and membrane tension, given by $\lambda=\sqrt{\kappa_b/\tau}$.  Comparing this `elastic decay length' with domain size indicates the set of possible domain morphologies.  If the domain size is on the order of, or smaller than, the elastic decay length, flat and dimpled morphologies arise; while domains larger than $\lambda$ generally give rise to the budded morphology \cite{Lipowsky1992,SensPRL2005,Harden2005,Taniguchi1996}.  This rule of thumb is based on the fact that bilayer deformations from dimpling are concentrated within a few elastic decay lengths of the phase boundary, while domain budding energetics are governed by basis shapes much larger than $\lambda$.

Dimpled domains are characterized by a dome-like shape with finite slope at the boundary between the two lipid phases, as shown in Figs.~\ref{fig1} and \ref{fig2}(b).  In previous work we found a distinct flat-to-dimpled transition, regulated by line tension and domain area \cite{Ursell2008}, where if all the elastic parameters are constant in time, the domain is either flat or dimpled, but cannot make transitions between those states ({\it i.e.}~there is no coexistence regime). Hence, domains below a critical size lie flat, and if all other membrane properties are constant, the only way domains transition from flat to dimpled is by coalescing to form larger domains that are above a critical size. An important outcome of domain dimpling is the emergence of a membrane-mediated repulsive interaction between domains that tends to inhibit coalescence.  Intuitively, the origin of this force is that dimples deform their surrounding membrane, but this deformation decays back to an unperturbed state within a few elastic decay lengths of the phase boundary. Two domains that are within a few elastic decay lengths of each other have overlapping deformed regions, and thus the total elastic free energy depends on the distance between the domains, leading to a net repulsion.  A relatively simple mechanical model and previous measurements show that the interaction between two dimpled domains can be approximated by the pair potential $V(r)\propto e^{-r/\lambda}$, where $r$ is the center-to-center separation between the domains \cite{Ursell2008}. 

This repulsive interaction arrests coalescence, and hence significantly affects the evolution of domain sizes in a phase-separated membrane.  For a simple physical model, in the absence of any interaction, domains would diffuse \cite{Veatch2007} and coalesce at a rate such that domain size scales as $t^{1/3}$ \cite{Bray2002,SeulPRL1994,Foret2005}.  This sets the time-scale for full phase separation on a typical giant unilamellar vesicle (GUV) at $\sim1$ minute (see Fig.~\ref{fig2}(c) and \cite{Veatch2003}). However, viewing repulsive interactions as an energetic barrier to domain growth, and given the measured barrier height of $\sim5k_BT$ (with $k_B=1.38\times10^{-23}\,J/\mbox{K}$ and $T\simeq300\,\mbox{K}$) \cite{Ursell2008}, the rate of domain coalescence slows by the Arrhenius factor $e^{-5}=0.007$.  A clear example of the difference in the rate of domain growth, with and without elastic interactions, is shown in Fig.~\ref{fig2}(d and c), respectively, and \cite{Yanagisawa2007}. Thus elastic repulsion is a plausible mechanism by which lipid lateral heterogeneity could be maintained on the hour-long time scale required for a cell to recycle (and hence partially homogenize) the plasma membrane \cite{Hansen1992}. Alternative schemes have been proposed that balance continuous rates of membrane recycling and domain coalescence to yield a stable domain size distribution \cite{SensPRL2005}. 

We examined the role these elastic interactions play in the spatial organization of lipid domains.  Given that all the domains mutually repel each other, as the areal density of domains increases, the arrangement of domains that minimizes the elastic free energy takes on distinctly hexagonal order, so as to maximize the separation between all domains.  Indeed, the arrangement of repulsive bodies on a sphere is a well studied problem in physics \cite{Bausch2003,Bowick2002,Erber1997}, and a dominant feature of such systems is the emergence of hexagonal and translational order.  We measured the strength of this organizing effect by tracking the thermal motion of dimples on the surface of GUVs, and calculating the radial distribution function (see `Materials and Methods').  For time-courses that have no coalescence events, the vesicle and its domains are in quasistatic equilibrium, thus the negative natural logarithm of the radial distribution function is a measure of the potential of mean force between domains.  Our previous theoretical and experimental work showed that the elastic interaction between domains at low areal density is well approximated by a pair potential of the form $V(r)\propto e^{-r/\lambda}$ \cite{Ursell2008}.  As the domain areal density increases, the domains adopt spatial orientations that maximize their mean spacing.  This can be understood in terms of the free energy of the entire group of domains.  If a domain deviates from this spatial arrangement, the sum of the elastic interaction energy with its neighboring domains increases, providing a mild restoring force to its original position.  Thus the potential of mean force develops energy wells, up to $\sim2\,k_BT$ in depth, that confine domains to a well-defined spacing, as shown in Fig.~\ref{fig3}(b-f).  It should be noted that such a restoring force can arise from the combined {\it pair} repulsion of the hexagonally-arranged neighboring domains, and does not necessarily mean that there are attractive, non-pairwise interactions.  

The mutual repulsion and resulting energetic confinement lead to an effective lattice constant that depends on domain size and packing density.  For example, domains may exhibit a well-defined spacing for first, second (Fig.~\ref{fig3}(b)), third (Fig.~\ref{fig3}(c-d)), and fourth (Fig.~\ref{fig3}(e-f)) nearest-neighbors, corresponding to a correlation in the position of domains over a few microns.  Interestingly, this means that by forming dimpled domains, the motion of individual lipids can be correlated on length scales up to $\sim10^4$ times larger than the size of an individual lipid.   As indicated by the exponential decay of these `ringing' potentials (see Fig.~\ref{fig3}(b-f)), the length scale over which this correlation in motion exists is limited by both the relatively low strength of pair repulsion (relative to $k_BT$) and the dispersion of domain sizes.  

In the picture that emerges, lipid domains exhibit a transition similar to condensation in a liquid--gas system.  At the lowest areal densities, the motion of domains is analogous to a `gas' of particles that occasionally have repulsive pairwise interactions.  As the domain areal density increases a `condensed' phase of domains emerges, identified by its translational and orientational order.  As the areal density of dimpled domains increases, the system exhibits three qualitative effects: i)  the lattice constant decreases, as it must, to accommodate more domains per unit area; ii)  the effective confinement grows stronger because the membrane in between the domains is more severely deformed by the closer packing; iii) hexagonal order clearly emerges, as shown by the characteristic peaks in the time-averaged Fourier transforms of domain positions in Fig.~\ref{fig3}(b-f)(inset).  The time-averaged Fourier transform is the arithmetic mean of the Fourier transforms of domain positions from each image in a data set, with the peaks corresponding to hexagonal order somewhat `smeared' by the rotational diffusion of the entire group of domains.  

To quantify the lattice constant and correlation length of interacting dimpled domains, we added a phenomenological correction term to the previously mentioned pair potential to account for interactions between multiple domains, such that the total potential of mean force has the form
\be{
V_{\mbox{\tiny fit}}(r)=a_1e^{-\frac{r}{\lambda_1}}+a_2e^{-\frac{r}{\lambda_2}}J_0\left(2\pi(r-r_o)/\lambda_3\right),
\label{fiteq}
}\ee
where $J_0$ is the $0^{\mbox{\tiny th}}$-order Bessel Function of the first kind, and $a_1$, $\lambda_1$, $a_2$, $\lambda_2$, $r_o$ and $\lambda_3$ are fit parameters.  While this equation offers little insight into the underlying physics of densely interacting domains, the description does an excellent job of capturing the observed features of interactions between multiple dimpled domains, as demonstrated in Fig.~\ref{fig3}.  Using this formula, we extracted the correlation length ($\lambda_2$) and lattice constant ($\lambda_3$), whose ratio, $\lambda_2/\lambda_3$, is a measure of the translational order in the system, which is shown to increase with domain density.  As domain areal density increases, the elastic free energy confines domains to adopt a well-defined mean spacing with hexagonal order.  Thus the motion of domains is correlated over multiple layers of neighboring domains, as shown in Fig.~\ref{fig3.25}. 

The dimpled domains that exhibit this behavior arise in situations where the tension is low and the elastic decay length is longer than the domain size. In the regime where the elastic decay length is short compared to domain size, `budded' domains emerge as a morphology with distinct transition rules and interactions.

\n\\
{\bf The Budding Transition\\}
Similar to the analysis of domain dimpling; bending, membrane tension, line tension, and domain size all play a role in the transition to a budded domain morphology.  Many energetic models have been proposed that describe morphological changes which result in budding and other more complex morphologies \cite{Harden2005,Seifert1991,Julicher1996,SensBJ2004,Gompper2001,Allain2006}.  One of the simplest models, and yet most reconcilable with experiment, is the `spherical' budding model. This model has its foundations in classical `sessile' droplet wetting theory \cite{Widom1995}, and posits that the domain is, at all times, a section of a sphere \cite{SensPRE2006,Lipowsky1992}.  We will recapitulate this model here, and explore some of its implications for our experiments.  This model ignores deformations near the phase boundary, and cannot capture the existence of the dimpled state, but is a reasonable model to employ in the regime where the elastic decay length is smaller than the domain size. 

The budding domain is characterized by a wrapping angle $\theta$, where $\theta=0$ corresponds to a flat domain and $\theta=\pi$ corresponds to the encapsulation of a small volume by a spherical bud, as shown in Fig.~\ref{fig1}(c) and ~\ref{fig3.5}(a).  The bending energy of a budding domain is calculated as a fraction of the bending energy of a sphere, given by
\begin{eqnarray}
G_{\mbox{\tiny bend}}&=&2\kappa_b\int\left(H-c_o\right)^2\dd \mathcal{A}\\
&=&8\pi\kappa_b\frac{\mathcal{A}}{4\pi R^2}\left(1-2c_oR\right)+C,\nonumber
\end{eqnarray}
where $H=1/R$ is the mean curvature, $R$ is the radius of curvature of the domain, $8\pi\kappa_b$ is the bending energy of a sphere, $\mathcal{A}$ is the domain area, and $c_o$ is the spontaneous curvature of the domain, which, for simplicity, we assume is zero.  A constant energy $C$, that does not depend on domain shape, is omitted.  As the domain becomes more spherical, the areal footprint of the domain shrinks, as shown in Fig.~\ref{fig3.5}(a), and work must be done against the applied membrane tension, given by
\be{
G_{\mbox{\tiny tens}}=-\tau\pi (R \sin{\theta})^2.
}\ee
The driving force for budding is the reduction of phase-boundary line tension, provided by
\be{
G_{\mbox{\tiny line}}=\gamma2\pi R\sin{\theta}.
}\ee
Finally, for all reasonable membrane tensions, the domain area is conserved during any change in morphology, and is given by
\be{
\mathcal{A}=2\pi R^2(1-\cos{\theta}).
}\ee
This constraint equation links $\mathcal{A}$ and $R$, allowing us to eliminate $R$ from the total free energy, $G=G_{\mbox{\tiny bend}}+G_{\mbox{\tiny tens}}+G_{\mbox{\tiny line}}$.  After some rearrangement, the total free energy can be written in a compact form,
\be{
G=4\pi\kappa_b[\,\,\underbrace{\chi\sqrt{\alpha}\sqrt{\frac{1+\cos\theta}{8\pi}}}_{\mbox{\tiny line tension}}-\underbrace{\alpha\frac{1+\cos\theta}{8\pi}}_{\mbox{\tiny membrane tension}}+\underbrace{1-\cos\theta}_{\mbox{\tiny bending}}\,\,],
}\ee
where the dimensionless area, $\alpha=\mathcal{A}/\lambda^2$, and dimensionless line tension, $\chi=\gamma\lambda/\kappa_b$, emerge as the regulators of domain budding.  The stable morphologies are those found at energy minima, given by $\partial G/\partial\theta=0$, with only flat ($\theta=0$) or budded ($\theta=\pi$) morphologies satisfying this equation (in the absence of spontaneous curvature). Figure \ref{fig3.5}(b) shows the free energy of budding as a function of wrapping angle $\theta$ and the line tension $\chi$.  From this plot, one can readily see that there are two special values of the line tension; the first, shown in blue, is where the free energy difference between the flat and budded states equals zero, but an energy barrier exists between them.  The second special value of line tension, shown in red, is where the energy barrier between flat and budded morphologies disappears, and budding becomes a spontaneous process.  This graphical analysis primes us to calculate the budding phase diagram. From the solutions for the energy minima, it can be shown that the phase diagram has three regions, as shown in Fig.~\ref{fig4}:  i) for certain values of $\alpha$ and $\chi$ both flat and budded domains are stable (coexistence), but flat domains have a lower elastic free energy; ii) in an adjacent region, both morphologies are stable (coexistence), but buds have a lower elastic free energy; iii) in the remaining region only buds are stable (single-phase).  The boundary between the regions of the phase diagram that have two stable morphologies (coexistence) versus one stable morphology (single-phase) is given by the inflection point $\left.(\partial^2 G/ \partial\theta^2 )\right|_{\theta=0}=0$, which defines the line tension
\be{
\chi_{\mbox{\tiny bud}}=8\sqrt{\frac{\pi}{\alpha}}+\sqrt{\frac{\alpha}{\pi}},
\label{chi_bud}
}\ee
above which only buds are stable, or alternatively stated, there is no energy barrier to the budding process (see Fig.~\ref{fig4}). Given that $\chi$ is a constant material parameter for constant tension, this equation specifies a size range over which spontaneous domain budding will occur,
\be{
\frac{\pi}{4}\left(\chi-\sqrt{\chi^2-32}\right)^2<\alpha<\frac{\pi}{4}\left(\chi+\sqrt{\chi^2-32}\right)^2.
\label{bud_range}
}\ee 
Thus budding, and in particular spontaneous budding, is a size-selective process that can only occur if $\chi>4\sqrt{2}$.  Membrane tension and line tension can be estimated by measuring size-selective spontaneous budding on the surface of a phase-separated vesicle. In a few instances, we were able to capture the onset of size-selective budding, though as a function of initial conditions and timing, this proves particularly difficult.  Sample temperature is a coarse knob that allows us to change the state of tension on the vesicle surface.  Though the exact value of the thermal area expansion coefficient for bilayers varies with composition, a good approximate value in the temperature range of interest is $c_{\mbox{\tiny exp}}\simeq5\times10^{-3}K^{-1}$ \cite{Needham1986,Needham1987,Sackmann1991}.  In Fig.~\ref{fig5}(a-c), the temperature is increased slightly (from $\sim18\,C$ to $\sim20\,C$, see `Materials and Methods'), increasing vesicle area by approximately 1\% while maintaining the enclosed volume, thus lowering the tension and driving the system into the spontaneous budding regime.  The average size of budding domains is $r=0.93\pm0.18\,\mu\mbox{m}$.  Using eqn.~\ref{bud_range}, and taking $\kappa_b=25\,k_BT$ as a nominal value for the bending modulus of a domain \cite{Baumgart2003,Evans2000,Chen1997,Ipsen2004}, we can solve the equations defined by the upper and lower bound to find the line tension and membrane tension.  From this analysis, we estimate $\tau\simeq2.4\times10^{-4}\,k_BT/\mbox{nm}^2$ and $\gamma\simeq0.45\,k_BT/\mbox{nm}$.  Using the tension and our assumption of bending modulus, we can also calculate the elastic decay length and dimensionless domain size to find $\lambda\simeq320\,\mbox{nm}$ and  $\alpha\simeq26$.  This membrane tension, which is within the range set by typical free vesicle \cite{Baumgart2003,Evans2000} and unstressed plasma membrane experiments \cite{Morris2001,Popescu2006}, sets the dimensionless domain area larger than one, and hence suggests that the spherical budding model is a good approximation. This estimate of line tension is consistent with previous measurements \cite{Baumgart2003,Tian2007}, and quantitatively matches results from our previous work \cite{Ursell2008}.

As a prelude to the calculation of allowed morphological transitions, we note that for the morphology of a domain to move from one region of the phase diagram to another, the domain must either change size via coalescence, or there must be a change in membrane tension which affects both $\alpha$ and $\chi$. On the phase diagram in Fig.~\ref{fig4}, horizontal lines would correspond to increasing domain area, and the dashed trajectories are increasing membrane tension with fixed domain area.  The key fact is that, except within a region very near the phase boundary, the free energy difference between the flat and the fully budded states is much larger than $k_BT$, as is the energy barrier between those states ({\it e.g.} Fig.~\ref{fig3.5}).  Thus, from an equilibrium statistical mechanics perspective, a budding domain can be approximated as a two state system, with the spontaneous budding regime included within the budded state. Thus, it makes sense to impose the thermodynamic requirement that the free energy difference between morphological states be negative for a transition to be allowed, {\it i.e.~}$G|_{\theta=\pi}-G|_{\theta=0}<0$ if going from flat to budded. This amounts to describing budding with a two-state model where
\be{
\Delta G_{\tiny f\rightarrow b}=G|_{\theta=\pi}-G|_{\theta=0}=\pi\kappa_b\left(\rho^2-2\chi\rho+8\right),
\label{twostateeq}
}\ee
and $\rho=\sqrt{\alpha/\pi}$ is the dimensionless domain radius.  Figures \ref{fig5}(a-c) and \ref{fig6}(b-d) show the two states of budding on the surface of phase-separated vesicles.  If we consider $\rho$, a measure of domain size, as an independent variable, then the single control-parameter $\chi$ dictates whether the thermodynamic condition $\Delta G_{\tiny f \rightarrow b}<0$ has been met for a particular domain size. If the dimensionless line tension is below the critical value $\chi_c=2\sqrt{2}$, defined by $\Delta G_{\tiny f \rightarrow b}=0$, the budding transition is forbidden for all domain sizes.  If $\chi>\chi_c$, budding is allowed, thought not necessarily spontaneous, within the size range given by
\be{
\rho=\chi\pm\sqrt{\chi^2-8},
\label{rho_size}
}\ee
as demonstrated in Fig.~\ref{fig5}(f). This size range always includes the range specified by eqn.~\ref{bud_range}, because spontaneous budding always has a negative free energy.

\n\\
{\bf The `Algebra' of Morphology\\}
With an understanding of the conditions under which a domain transitions from flat to dimpled \cite{Ursell2008}, and dimpled to budded, we are in position to calculate the change in free energy when domains of different morphologies coalesce.  On a short enough time-scale, coalescence only occurs between two domains at a time, and hence we can think of the coarsening behavior of a phase-separated membrane as many such binary coalescence events happening in succession.  The purpose of this section is to begin to build a framework for understanding how domain morphology and coalescence work in concert to affect the morphological evolution of a phase-separated membrane.  In particular, we calculate the allowed, resultant morphology when two domains, each of a distinct morphology, coalesce.  The change in free energy associated with a change in domain morphology, from flat to dimpled, or dimpled to budded, is much greater than $k_BT$, and hence, like the budding analysis of the previous section, for a particular transition to be allowed, we demand that the change in free energy be negative.  Furthermore, the large reduction in line energy upon coalescence (compared to $k_BT$) means that, in general, coalescence is irreversible, and hence after each coalescence event the system is presumed to be in a unique quasistatic equilibrium state, with a unique membrane tension.  The use of these transitions rules must then be considered in the context of these unique states, that is, transitions involving domains of a particular size that were allowed before a coalescence event might be prohibited afterward, or {\it vica versa}.

Let us denote transitions that involve flat domains with the letter $f$, dimpled domains with the letter $d$, and budded domains with letter $b$, such that, for instance, a flat domain coalescing with a dimpled domain to yield a budded domain would be denoted by $fd\rightarrow b$.  There are six possible binary coalescence events:  $ff$, $fd$, $fb$, $dd$, $db$, and $bb$; each resulting in a single domain of either $f$, $d$ or $b$ morphology.  Thus at the onset, there are a total of 18 possible morphological transitions, however, not all of them are thermodynamically allowed.  Specifically, any domain whose size is greater than the critical size required for dimpling cannot adopt a flat morphology as there is no flat-dimple coexistence, hence only the $ff\rightarrow f$ transition can end with an $f$ domain (see Fig.~\ref{fig7}(a)).  This eliminates five of the six possibilities that end with an $f$ domain.  The only other transition that can be eliminated immediately is $ff\rightarrow b$, because the coalescence of two flat domains must first go through the dimpled state.

This leaves twelve possible morphological transitions, as shown in Fig.~\ref{fig7}(a-$\ell$).  For simplicity we will assume the domains have no spontaneous curvature (though this is straightforward to incorporate \cite{Lipowsky1992,Gompper2003}).  The free energy change associated with each of these twelve transitions is calculated by knowing the free energy change associated with three simpler transitions, namely the $f\rightarrow d$, $f\rightarrow b$, and $ff\rightarrow f$ transitions.  Of these, the $f\rightarrow b$ transition free energy was discussed in the previous section, and the $f\rightarrow d$ transition free energy is a complicated function discussed at length in \cite{Ursell2008}, though we note the important fact that $\Delta G_{\tiny f\rightarrow d}(\alpha_1+\alpha_2)<\Delta G_{\tiny f\rightarrow d}(\alpha_1)+\Delta G_{\tiny f\rightarrow d}(\alpha_2)$ if both domains are above the critical size for dimpling, or in words, the free energy of domain dimpling as a function of domain area grows faster than linearly.  The scheme we are about to build is a valid frame work for understanding energy based transitions because we know that changes in free energy, when moving along a reaction coordinate, are additive.

The transition of two flat domains coalescing to yield another flat domain is the most fundamental transition, as shown in Fig.~\ref{fig7}, and can be calculated as the difference in the line tension energy between the initial and final states given by
\be{
\Delta G_{\tiny ff\rightarrow f}(\alpha_1,\alpha_2)=2\sqrt{\pi}\kappa_b\chi\left[\sqrt{\alpha_1+\alpha_2}-\sqrt{\alpha_1}-\sqrt{\alpha_2}\right],
\label{trans_a}
}\ee
where $\alpha_1$ and $\alpha_2$ are the dimensionless areas of the two domains and $\Delta G_{\tiny ff\rightarrow f}(\alpha_1,\alpha_2)<-k_BT$ for all domain areas of one lipid or more. This situation, depicted by Fig.~\ref{fig7}(a) and shown experimentally in Fig.~\ref{fig2}(c), is encountered at high membrane tension, when domains are too small to dimple before and after coalescence. 

Using the three basic transitions, we now address the remaining eleven transitions in detail.  The next transition we consider is two flat domains, each too small to dimple on their own, coalescing to form a domain large enough to dimple, as depicted in Fig.~\ref{fig7}(b).  The transition free energy is given by
\be{
\Delta G_{\tiny ff\rightarrow d}(\alpha_1,\alpha_2)=\Delta G_{\tiny ff\rightarrow f}(\alpha_1,\alpha_2)+\Delta G_{\tiny f\rightarrow d}(\alpha_1+\alpha_2),
\label{trans_b}
}\ee
and is negative as long as $\alpha_1+\alpha_2$ is greater than the critical area required for dimpling.

The next transition is a flat and dimpled domain coalescing to form a dimpled domain, as depicted in Fig.~\ref{fig7}(c).  The transition free energy is given by
\be{
\Delta G_{\tiny fd\rightarrow d}(\alpha_1,\alpha_2)=\Delta G_{\tiny ff\rightarrow d}(\alpha_1,\alpha_2)-\Delta G_{\tiny f\rightarrow d}(\alpha_2).
\label{trans_c}
}\ee
No definitive statement about the resultant morphology after coalescence of a flat and dimpled domain can be made, because the free energy of this transition must be compared to the closely related transition of a flat and dimpled domain coalescing to form a budded domain, to determine which has a greater reduction in free energy.   This related transition, depicted in Fig.~\ref{fig7}(h), has the transition free energy
\begin{eqnarray}
\label{trans_d}
\Delta G_{\tiny fd\rightarrow b}(\alpha_1,\alpha_2)=\Delta G_{\tiny ff\rightarrow f}(\alpha_1,\alpha_2)-\Delta G_{\tiny f\rightarrow d}(\alpha_2)\\
+\Delta G_{\tiny f\rightarrow b}(\alpha_1+\alpha_2)\nonumber.
\end{eqnarray}
Which of these two transitions, $fd\rightarrow d$ or $fd\rightarrow b$, dominates depends on which has a greater reduction in free energy.  Comparing eqns.~\ref{trans_c} and \ref{trans_d}, asking which transition has the greater reduction in free energy is simply asking whether $\Delta G_{\tiny f\rightarrow d}(\alpha_1+\alpha_2)>\Delta G_{\tiny f\rightarrow b}(\alpha_1+\alpha_2)$ or {\it vice versa}.  This energy balance between the $f\rightarrow d$ and $f\rightarrow b$ transitions determines the outcome of all of the subsequent binary transitions as well, though we will rigorously show this for the remaining cases.  Because this energetic comparison crops up so often, we will simply refer to it as the `bud-dimple energy balance.'

The next transition is two dimpled domains coalescing to yield a dimpled domain, as depicted in Fig.~\ref{fig7}(d) and shown in Fig.~\ref{fig6}(a).  The transition free energy is given by
\be{
\Delta G_{\tiny dd\rightarrow d}(\alpha_1,\alpha_2)=\Delta G_{\tiny fd\rightarrow d}(\alpha_1,\alpha_2)-\Delta G_{\tiny f\rightarrow d}(\alpha_2).
\label{trans_e}
}\ee
Again, we must consider a closely related transition, namely the coalescence of two dimpled domains yielding a budded domain, as depicted in Fig.~\ref{fig7}(i), with transition free energy
\begin{eqnarray}
\Delta G_{\tiny dd\rightarrow b}(\alpha_1,\alpha_2)=\Delta G_{\tiny dd\rightarrow d}(\alpha_1,\alpha_2)-\Delta G_{\tiny f\rightarrow d}(\alpha_1+\alpha_2)\\
+\Delta G_{\tiny f\rightarrow b}(\alpha_1+\alpha_2)\nonumber.
\label{trans_f}
\end{eqnarray}
Comparing these two related transitions, $dd\rightarrow d$ and $dd\rightarrow b$, we see that the dominant transition is determined by the bud-dimple energy balance.

The next transition is a flat and a budded domain coalescing to form a dimpled domain, as depicted in Fig.~\ref{fig7}(g).  The transition free energy is given by
\be{
\Delta G_{\tiny fb\rightarrow d}(\alpha_1,\alpha_2)=\Delta G_{\tiny ff\rightarrow d}(\alpha_1,\alpha_2)-\Delta G_{\tiny f\rightarrow b}(\alpha_2).
\label{trans_g}
}\ee 
The related transition, where a flat and budded domain coalesce to form a budded domain, as depicted in Fig.~\ref{fig7}($\ell$), has the transition free energy
\begin{eqnarray}
\Delta G_{\tiny fb\rightarrow b}(\alpha_1,\alpha_2)=\Delta G_{\tiny fb\rightarrow d}(\alpha_1,\alpha_2)-\Delta G_{\tiny f\rightarrow d}(\alpha_1+\alpha_2)\\
+\Delta G_{\tiny f\rightarrow b}(\alpha_1+\alpha_2)\nonumber.
\label{trans_h}
\end{eqnarray}
Comparing these two related transitions, $fb\rightarrow d$ and $fb\rightarrow b$, we see that the dominant transition is determined by the bud-dimple energy balance.

The next transition is a budded and a dimpled domain coalescing to form a budded domain, as depicted in Fig.~\ref{fig7}(j) and shown in Fig.~\ref{fig6}(c-e)(green arrows).  The transition free energy is given by
\be{
\Delta G_{\tiny bd\rightarrow b}(\alpha_1,\alpha_2)=\Delta G_{\tiny fd\rightarrow b}(\alpha_1,\alpha_2)-\Delta G_{\tiny f\rightarrow b}(\alpha_2).
\label{trans_i}
}\ee 
The related transition, where a budded and dimpled domain coalesce to form a dimpled domain, as depicted in Fig.~\ref{fig7}(e), and shown in Fig.~\ref{fig6}(c-e)(yellow arrows), has the transition free energy
\begin{eqnarray}
\Delta G_{\tiny bd\rightarrow d}(\alpha_1,\alpha_2)=\Delta G_{\tiny bd\rightarrow b}(\alpha_1,\alpha_2)-\Delta G_{\tiny f\rightarrow b}(\alpha_1+\alpha_2)\\
+\Delta G_{\tiny f\rightarrow d}(\alpha_1+\alpha_2)\nonumber.
\label{trans_j}
\end{eqnarray}
Comparing these two related transitions, $bd\rightarrow b$ and $bd\rightarrow d$, we see that the dominant transition is determined by the bud-dimple energy balance.

The last set of transitions is when two buds coalesce to form a larger bud, depicted in Fig.~\ref{fig7}(k), with transition energy
\be{
\Delta G_{\tiny bb\rightarrow b}(\alpha_1,\alpha_2)=\Delta G_{\tiny fb\rightarrow b}(\alpha_1,\alpha_2)-\Delta G_{\tiny f\rightarrow b}(\alpha_2),
\label{trans_k}
}\ee 
and when two buds coalesce to form a dimple, depicted in Fig.~\ref{fig7}(f) with transition energy
\be{
\Delta G_{\tiny bb\rightarrow d}(\alpha_1,\alpha_2)=\Delta G_{\tiny fb\rightarrow d}(\alpha_1,\alpha_2)-\Delta G_{\tiny f\rightarrow b}(\alpha_2).
\label{trans_k}
}\ee
Comparing these two related transitions, $bb\rightarrow b$ and $bb\rightarrow d$, we see that the dominant transition is determined by the bud-dimple energy balance.

Given the importance of the bud-dimple energy balance in determining the morphology resulting from a coalescence event, we note that if the resulting domain area is outside the range specified by eqn.~\ref{rho_size}, but still larger than the critical size required for dimpling, the dimpled morphology dominates because the free energy change of budding is positive outside that range.  Within this size range, selecting the dominant behavior is more subtle, and depends on the resultant domain size, material properties and tension.  For this reason, until experimental methods are devised that can track the detailed three dimensional morphology of a phase separated vesicle ({\it i.e.} the positions and sizes of all domains and the membrane tension), the set of transition rules discussed in this section will remain largely an interpretive tool, useful for understanding the set of possible transitions and resultant morphologies, as well as their underlying physics, but difficult to quantitatively apply to experiment.

We speculate that the kinetics of these coalescence transitions are either relatively fast, when diffusion is the limiting time scale, as might be the case in the transitions shown in Fig.~\ref{fig7}(a-c,e,g,h,j,$\ell$), or relatively slow, limited by elastic interactions (Fig.~\ref{fig7}(d,i)) or steric hindrance (Fig.~\ref{fig7}(f,k)).  From the viewpoint of coarsening of a two-phase fluid, these transitions represent new coarsening mechanisms that are linked to morphology, and likely have profound effects on the kinetics of phase separation, as demonstrated by the fact that coalescence of dimpled domains is inhibited by an energetic barrier. Additionally, these transitions suggest interesting biological possibilities.  For instance, a small volume can be encapsulated at a particular location, as a dimple transitions to a bud.  The enclosed volume can then diffuse to other regions of the membrane, and either engulf more volume (see Fig.~\ref{fig7}(i)) or deposit its contents at the site of another domain (see Fig.~\ref{fig7}(j)).  In fact, both of these scenarios play out in Fig.~\ref{fig6}(c-e).  Furthermore, it is possible that careful control of membrane tension \cite{Sheetz1999} could regulate how large a volume is enclosed, and to which other domains a bud will coalesce and deposit its contents.  This has implications for the size-selectivity of endo- and exo-cytosis where membrane invagination and fusion occur, as well as regulation of plasma membrane tension \cite{Sheetz1999}. 


\n\\
{\bf Discussion\\}
The transition free energies calculated in the previous section have the intuitively pleasing feature of being a sum of three simple basis transitions ($f\rightarrow d$, $f\rightarrow b$ and $ff\rightarrow f$) . However, this type of analysis is limited by the fact that it only admits flat, dimpled and budded as valid morphologies.  More general theories and computational models can (and have been) constructed that attempt to describe all possible shapes of a domain from precisely flat to fully budded, and other more complex morphologies \cite{Taniguchi1996,Seifert1991,Julicher1996,Gompper2001,LaradjiPRL2004,Laradji2006,Hong2007}.  Our level of experimental sophistication is commensurate with the simplicity of the analysis employed in the previous section.  Conceptually, our model simplifies analysis by reducing domain morphology to one of three classes of shapes, at the cost of excluding other possible morphologies.  Though overall an experimental minority, domain-induced tube formation was the most common of these more exotic morphologies.  Normally, thin lipid tubes are drawn out by external force \cite{Dogterom2005,Prost2002,Powers2002}.  However, in a few instances we observed domains that spontaneously collapse and nucleate a tube that rapidly grows many times longer than its persistence length, as demonstrated in Fig.~\ref{fig8}.  Oddly, the nucleating domain is of one lipid phase, but the tube continues to grow from the other, majority phase by a currently unknown mechanism.

In addition to limiting the class of possible morphologies, our analysis of morphological transitions also employs the simplification that membrane tension is constant during a morphological transition.  In reality, our experiments take place on a spherical topology with constrained volume and surface area, such that this approximation has a range of validity.  If the membrane area required to complete a morphological transition is small compared to the total vesicle area (see \cite{Ursell2008} for details) the change in membrane tension will be small. However, morphological transitions that require relatively large areas can result in significant changes to membrane tension, invalidating the constant tension approximation.  Although, at times this can be an advantageous feature of our experimental system, for instance, when fairly small changes in vesicle area (on the order of 1\%) can reduce the tension enough to cause spontaneous budding, as we showed earlier in this work.

In addition to the limitations mentioned above, our experiments have a number of subtle complications.  Notably, the task of measuring the motion and size of lipid domains is complicated by the fact that the spherical curvature of the vesicle slightly distorts measurements of distance and size.  Additionally, the motion of domains is confined to lie in a circle defined by a combination of vesicle size and depth of field of the microscope objective.  We developed schemes to correct for these issues, as discussed in detail in the supplementary information of our previous work \cite{Ursell2008}.




\n\\
{\bf Summary\\}
Using a model multi-component membrane, we explored how the interplay between composition and morphology leads to elastic forces that spatially organize domains and significantly impact coalescence kinetics.  We expanded upon mechanical models that incorporate bending stiffness, membrane tension, phase boundary line tension, and domain size to show that domains can adopt (at least) three distinct morphologies: flat, dimpled and budded.  We showed that dimpled domains exhibit measurable translational and orientational order as a function of increasing domain areal density \cite{Bausch2003,Bowick2002,Erber1997}.  Using a spherical budding model, we showed that the transition to a budded state is a domain size selective process, from which one can estimate the membrane tension, line tension, and elastic decay length of a phase separated membrane. Additionally, we found that the large energy scales associated with changes in domain morphology allow us to define morphological transition rules, where domain size and membrane tension are likely the key parameters that regulate the morphological transitions.

In the context of our understanding of the physics of phase separation the elastic forces between dimpled domains that arrest coalescence, and the morphological transitions between flat, dimpled and budded domains, constitute new mechanisms that govern spatial organization of domains and the temporal evolution of domain sizes.  For cellular membranes, we speculate that the elastic forces and morphological transitions can be controlled via careful regulation of membrane tension \cite{Sheetz1999}, and our work suggests intriguing possibilities for how small volumes can be encapsulated, moved, and released in a phase-separated membrane.


\n\\
{\bf Materials and Methods\\} 
Giant unilamellar vesicles (GUVs) were prepared from a mixture of DOPC (1,2-Dioleoyl-sn-Glycero-3-Phosphocholine), DPPC (1,2-Dipalmitoyl-sn-Glycero-3-Phosphocholine) and cholesterol (Avanti Polar Lipids) (25:55:20/molar) that exhibits liquid-liquid phase coexistence \cite{Veatch2003}.  Fluorescence contrast between the two lipid phases is provided by the rhodamine head-group labeled lipids: DOPE (1,2-Dioleoyl-sn-Glycero-3-Phosphoethanolamine-N- (Lissamine Rhodamine B Sulfonyl)) or DPPE (1,2-Dipalmitoyl-sn-Glycero-3-Phosphoethanolamine-N- (Lissamine Rhodamine B Sulfonyl)), at a molar fraction of $\sim0.005$.  The leaflet compositions are presumed symmetric and hence $c_o=0$.  

GUVs were formed via electroformation \cite{Veatch2003,Angelova1992}.  Briefly, $3-4\,\mu\mbox{g}$ of lipid in chloroform were deposited on an indium-tin oxide coated slide and dessicated for $\sim2\,\mbox{hrs}$ to remove excess solvent.  The film was then hydrated with a $100\,\mbox{mM}$ sucrose solution and heated to $\sim50\,\mbox{C}$ to be above the miscibility transition temperature.  An alternating electric field was applied; $10\,\mbox{Hz}$ for 120 minutes, $2\,\mbox{Hz}$ for 50 minutes, at $\sim500\,\mbox{Volts/m}$ over $\sim2\,\mbox{mm}$.  Low membrane tensions were initially achieved by careful osmolar balancing with sucrose ($\sim100\,\mbox{mM}$) inside the vesicles, and glucose ($\sim100-108\,\mbox{mM}$) outside. Using a custom built temperature control stage, the {\it in situ} membrane tension was coarsely controlled by adjusting the temperature a few degrees \cite{Needham1987,Sackmann1991}.

Domains were induced by a temperature quench and imaged using standard TRITC epi-fluorescence microscopy at 80x magnification with a cooled (-30 C) CCD camera (Roper Scientific, $6.7\times6.7\,\mu\mbox{m}^2$ per pixel, 20 MHz digitization).  Images were taken from the top or bottom of a GUV where the surface metric is approximately flat.  Data sets contained $\sim500-1500$ frames collected at 10-20 Hz with a varying number of domains (usually $>10$).  The frame rate was chosen to minimize exposure-time blurring of the domains, while allowing sufficiently large diffusive domain motion.   Software was written to track the position of each well-resolved domain and calculate the radial distribution function.  The raw radial distribution function was corrected for the fictitious confining potential of the circular geometry. The negative natural logarithm of the radial distribution function is the potential of mean force plus a constant, as shown in Fig.~\ref{fig3}.  Detailed explanations of these concepts can be found in the supplementary information for \cite{Ursell2008}.

Morphological transitions were induced by quenching homogeneous vesicles below the de-mixing temperature and observing those that had many micron-sized domains.  Without precise control of membrane tension or the exact initial conditions ({\it i.e.} the exact number and size distribution of domains) many vesicles had to be sampled to see transitions.  Often, a slight increase in temperature ($\sim2$C) was used to increase the available membrane area, and hence decrease the membrane tension enough to induce morphological transitions.\\

\footnotesize{We thank Jennifer Hsiao for help with experiments, and Kerwyn Huang, Ben Freund and Pierre Sens for stimulating discussion and comments.  TSU and RP acknowledge the support of the National Science Foundation award No.~CMS-0301657, NSF CIMMS award No.~ACI-0204932, NIRT award No.~CMS-0404031 and the National Institutes of Health award No. R01 GM084211 and the Director's Pioneer Award.}

{\tiny

}
}

\onecolumn
\begin{figure}
\begin{center}
\includegraphics[width=7in]{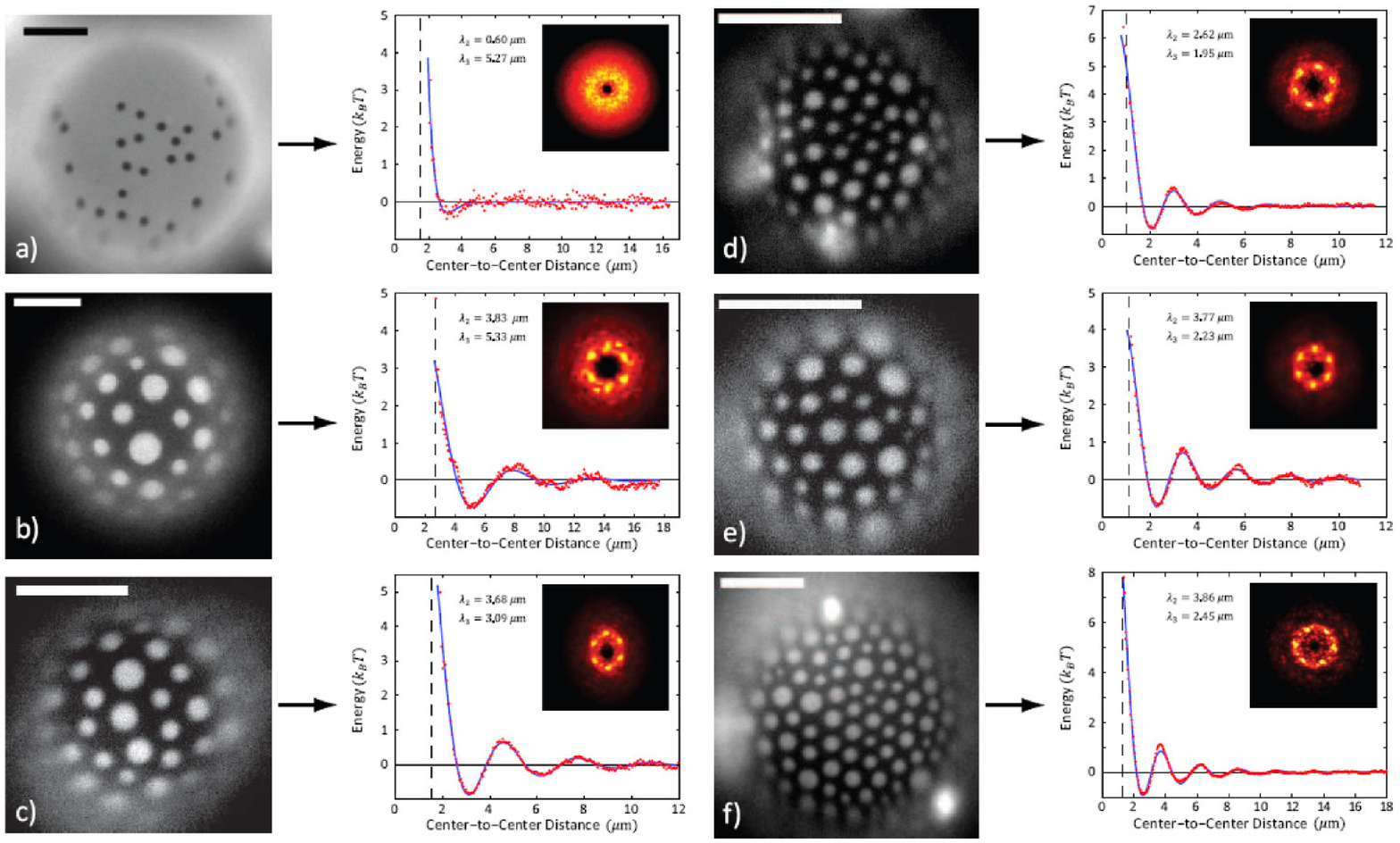}
\caption{\footnotesize{Interactions of lipid domains as areal density increases.  Left: snapshots of dimples on the surface of GUVs.  Right: corresponding potentials of mean force. a) At low areal density, interactions are almost purely repulsive, and there is no translational or orientational order -- the domains are in a state analogous to a gas of particles. b-f) At higher areal density, domains `condense' into a state where each domain is repelled by its neighboring domains, giving rise to energy wells that define a lattice constant and hence translational order. The decay envelope of these `ringing' potentials indicates the length-scale over which the motion of domains is correlated.  In all plots, the blue line indicates the fit to eqn.~\ref{fiteq}, where $\lambda_2$ is the order-correlation length and $\lambda_3$ is the effective lattice constant.  The dashed vertical lines are the approximate minimum center-to-center distance between domains as determined by domain size measurement (a) or one half the lattice constant (b-f).  Insets: Time-averaged Fourier transforms, showing that mutually repulsive elastic interactions lead to (thermally smeared) hexagonal order, except in (a) where the density is too low to order the domains. Scale bars are $10\,\mu\mbox{m}$.}}
\label{fig3}
\end{center}
\end{figure}
\twocolumn

\begin{figure}
\begin{center}
\includegraphics[width=2.5in]{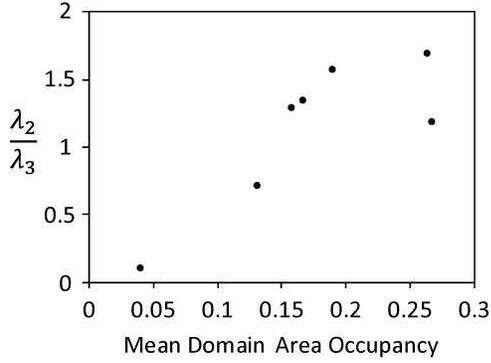}
\caption{\footnotesize{Increase in correlated motion of dimpled domains as a function of area occupancy.  This plot shows the ratio of the order-correlation length ($\lambda_2$) over the lattice constant ($\lambda_3$) for the vesicles in Fig.~\ref{fig3} (and one additional vesicle) as a function of the total area taken up by the domains divided by the total measurable vesicle area.  The ratio $\lambda_2/\lambda_3$ quantifies how many nearest neighbor domains ({\it i.e.}~1st, 2nd, etc.) exhibit strongly correlated motion.}}
\label{fig3.25}
\end{center}
\end{figure}

\begin{figure}
\begin{center}
\includegraphics[width=3.1in]{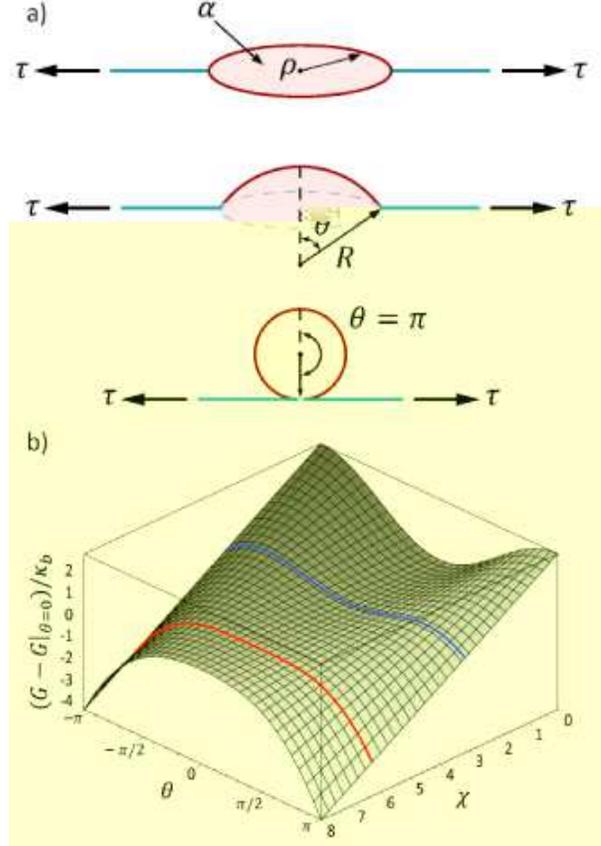}
\caption{\footnotesize{Shapes and energies of domain budding. a)  A schematic of domain shape going from a flat domain, with area $\alpha$ and flat radius $\rho$, through a dome shape with wrapping angle $\theta$ and radius of curvature $R$, to a fully budded state, with an applied tension $\tau$. b) The free energy of a budding domain as a function of line tension ($\chi$) and wrapping angle ($\theta$) for domain size $\alpha=10$.  At low line tension (before the blue line), both flat and budded morphologies are stable, but the flat state has a lower elastic free energy and there is an energy barrier between the two stable states.  At the blue line, the free energy difference between flat and budded is zero.  Between the blue and red lines, both morphologies are stable, but the budded state has a lower elastic free energy.  Finally, for line tensions above the red line, the energy barrier disappears and budding is a spontaneous process.}}
\label{fig3.5}
\end{center}
\end{figure}

\begin{figure}
\begin{center}
\includegraphics[width=3.1in]{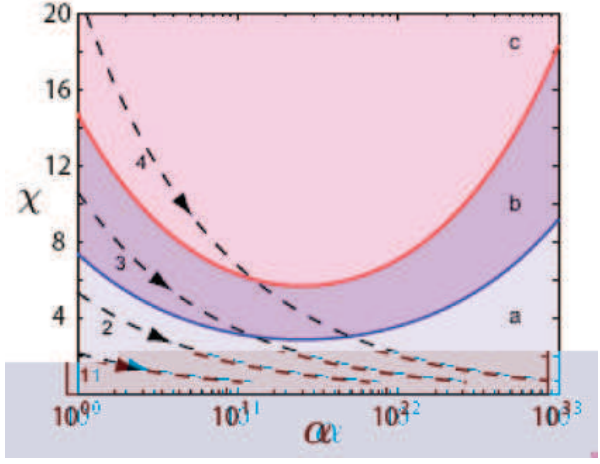}
\caption{\footnotesize{Equilibrium phase diagram for domain budding as a function of dimensionless domain area and line tension.   In region `a' flat and budded domains coexist, with flat domains at lower free energy.  In region `b' flat and budded domains coexist, with budded domains at lower free energy.  In region `c' only a single, budded phase is stable. The line separating regions `a' and `b' is given by $\chi_{\mbox{\tiny bud}}/2$ and between regions `b' and `c' by $\chi_{\mbox{\tiny bud}}$ (eqn.~\ref{chi_bud}).  Dashed lines are trajectories of increasing membrane tension (as indicated by the arrows) at constant domain area.  In all four trajectories $\gamma=0.3\,k_BT/\mbox{nm}$, $\kappa_b=25\,k_BT$ and tension is varied from $\tau=10^{-5}-10^{-2}\,k_BT/\mbox{nm}^2$; the domain areas are $\mathcal{A}_1=\pi(100\mbox{nm})^2$, $\mathcal{A}_2=\pi(250\mbox{nm})^2$, $\mathcal{A}_3=\pi(500\mbox{nm})^2$, and $\mathcal{A}_4=\pi(1000\mbox{nm})^2$.}}
\label{fig4}
\end{center}
\end{figure}

\begin{figure}
\begin{center}
\includegraphics[width=3.1in]{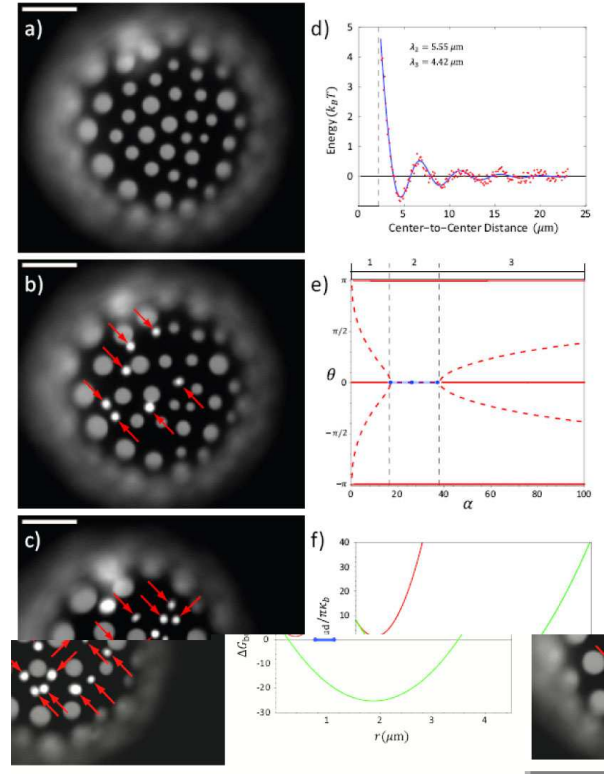}
\caption{\footnotesize{Size-selective domain budding.   a) Dimples on the surface of a GUV are initially arranged by their repulsive interactions.  b) and c) A slight increase in temperature decreases membrane tension, causing smaller domains to spontaneously bud (marked with red arrows) and wander freely on the vesicle surface, while larger domains remain dimpled.  The mean size of budding domains is $r\simeq0.93\pm0.18\,\mu\mbox{m}$ from which we estimate a line tension of $\gamma\simeq0.45\,k_BT/\mbox{nm}$.  d) Plot of the potential of mean force between the dimpled domains in (a), moments before inducing spontaneous budding.  e) Budding stability diagram, showing solutions to $\partial G/\partial\theta=0$ with $\gamma=0.45\,k_BT/\mbox{nm}$.  Solid red lines are stable solutions at energy minima; dashed red lines are unstable solutions on the energy barriers.   Regions 1 and 3 are coexistence regimes, while region 2 is a spontaneous budding regime, only stable at $|\theta|=\pi$.  The blue dots indicate domain areas with radii $r\simeq0.93,0.93\pm0.18\,\mu\mbox{m}$. f) The red curve shows the free energy of budding at $\tau=1.2\times10^{-3}\,k_BT/\mbox{nm}^2$, which is greater than zero for all domain sizes, and hence all domains would remain flat/dimpled. The green curve shows the free energy of budding at $\tau=2.4\times10^{-4}\,k_BT/\mbox{nm}^2$.  Domains with radius $r\simeq0.25-3.5\,\mu\mbox{m}$ have a negative free energy of budding, all other sizes remain flat/dimpled.  Most domains within this size range must still overcome an energy barrier to bud, but for a small range of domain sizes ($r\simeq0.75-1.11\,\mu\mbox{m}$), indicated by the blue line segment, budding is a spontaneous process.   The energies are calculated using $\kappa_b=25\,k_BT$ and $\gamma\simeq0.45k_BT/\mbox{nm}$.  In (a-c) scale bars are $10\,\mu\mbox{m}$.}}
\label{fig5}
\end{center}
\end{figure}

\onecolumn
\begin{figure}
\begin{center}
\includegraphics[width=7in]{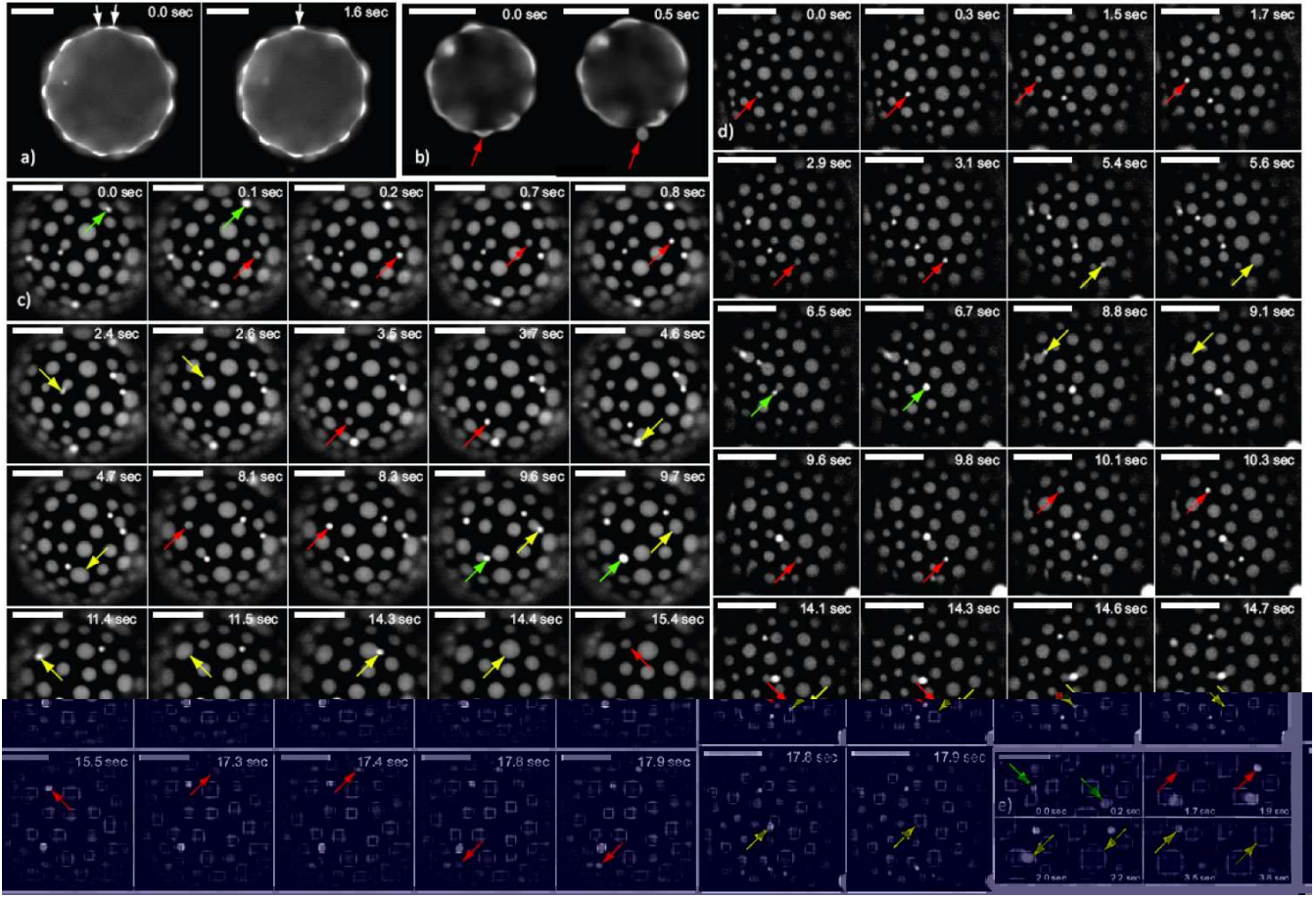}
\caption{\footnotesize{Gallery of morphological transitions.  a) Two dimpled domains (indicated by the white arrows) interact on the surface of vesicle, eventually coalescing to yield a larger dimpled domain (see Fig.~\ref{fig7}(d)). b) An equatorial view of a dimple-to-bud transition (indicated by the red arrows).  c-e)  Time courses of multiple types of morphological transitions.  Arrows are color-coded and point to before and after each transition:  red arrows indicate a dimple to bud transition, green arrows indicate a bud engulfing a dimple to form a larger bud (see Fig.~\ref{fig7}(j)), and yellow arrows indicate a bud recombining with a larger dimpled domain (see Fig.~\ref{fig7}(e)).  Using video microscopy, we can put an upper bound on the time scale of the $d\rightarrow b$, $db\rightarrow d$ and $db\rightarrow b$ transitions at $\sim200\pm80\,\mbox{ms}$, $\sim160\pm70\,\mbox{ms}$ and $\sim210\pm70\,\mbox{ms}$, respectively. Scale bars are $10\,\mu\mbox{m}$.}}
\label{fig6}
\end{center}
\end{figure}

\begin{figure}
\begin{center}
\includegraphics[width=6.5in]{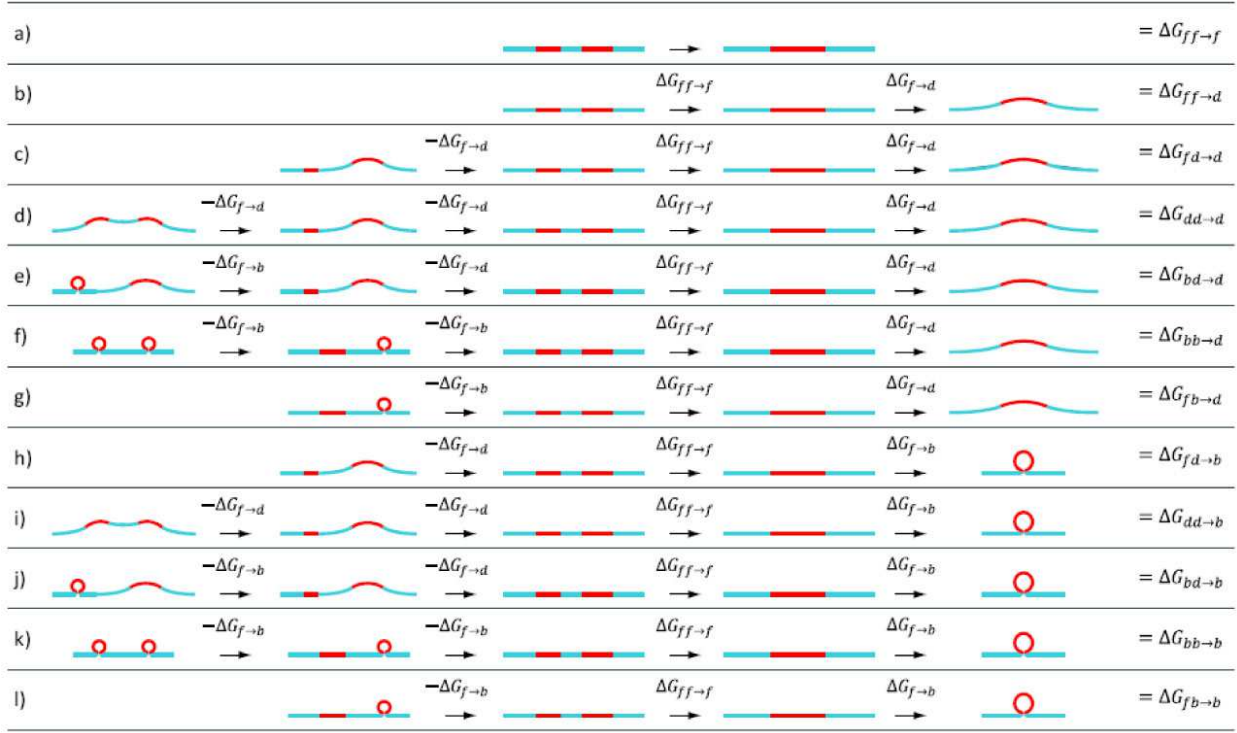}
\caption{\footnotesize{Algebra of Morphology.  All rows show additive free energies of transition from an initial state on the left to a final state on the right.  a) Two flat domains coalesce via diffusion, yielding a flat domain. b)  Domains, too small to dimple, coalesce to attain a size capable of dimpling. c) A dimple and a domain too small to dimple coalesce to yield a larger dimpled domain.  d) Two interacting dimples coalescence, yielding a larger dimpled domain. e) A dimple and a bud coalesce to yield a larger dimpled domain. f) Two buds coalesce to yield a dimpled domain.  g) A flat domain coalesces with a bud, yielding a dimpled domain. h) A flat domain coalesces with a dimpled domain to yield a bud.  i) Two interacting dimples coalescence, yielding a bud. j) A bud coalesces with a dimple, yielding a larger budded domain. k) Two budded domains coalesce to form a larger budded domain. l) A flat domain coalesces with a bud to yield a larger budded domain.}}
\label{fig7}
\end{center}
\end{figure}
\twocolumn

\begin{figure}
\begin{center}
\includegraphics[width=3in]{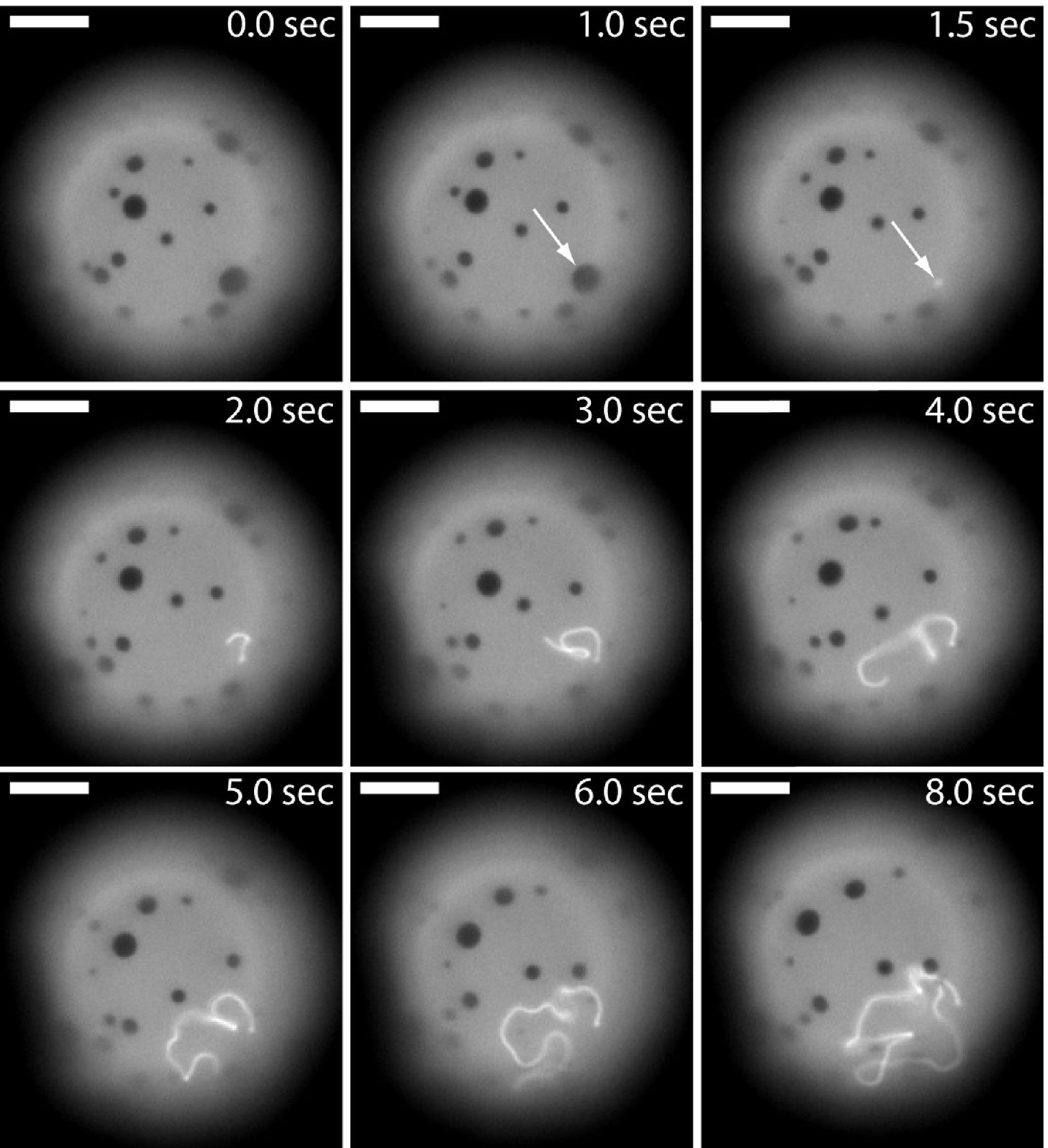}
\caption{\footnotesize{Domain-nucleated spontaneous tube formation.  A time series of spontaneous tube formation, nucleated from a domain (as indicated by the white arrow).  This relatively uncommon morphology is not explained within the context of our simple model.  The lipid tube (bright) is many times longer than its persistence length, yet perplexingly, grows from the tube {\it tip}. With limited optical resolution, we estimate the tube diameter to be $\leq500\,\mbox{nm}$.  The scale bar is $10\,\mu\mbox{m}$.}}
\label{fig8}
\end{center}
\end{figure}

\end{document}